\documentclass[Review,doublespace,times,sagev]{sagej}

\usepackage{moreverb,url,enumitem,bbm}
\usepackage[ruled,vlined]{algorithm2e}
\usepackage[colorlinks,bookmarksopen,bookmarksnumbered,citecolor=red,urlcolor=red]{hyperref}
\usepackage{pdflscape}
\usepackage{enumitem}
\usepackage{amsthm}
\usepackage{makecell}
\usepackage{multirow}

\newcommand\BibTeX{{\rmfamily B\kern-.05em \textsc{i\kern-.025em b}\kern-.08em
T\kern-.1667em\lower.7ex\hbox{E}\kern-.125emX}}

\def\volumeyear{2016}

\setcitestyle{super}

\theoremstyle{remark}

\newtheorem{lemma}{Lemma}
\newtheorem{proposition}{Proposition}
\newtheorem{remark}{Remark}

\begin{document}

\runninghead{Shoji, Yokota, and Herrmann}

\title{Stratified Randomization Incorporating Historical Control Data}

\author{Masatoshi Shoji\affilnum{1}, Isao Yokota\affilnum{1}, Carolin Herrmann\affilnum{2}} % to be revised

\affiliation{
\affilnum{1}Department of Biostatistics, Graduate School of Medicine, Hokkaido University, Sapporo, Japan.\\
\affilnum{2}Mathematical Institute and Center for Digital Medicine, Faculty of Mathematics and Natural Sciences, Heinrich Heine University Düsseldorf, Universitätsstraße 1, 40225 Düsseldorf, Germany}

\corrauth{Isao Yokota, Department of Biostatistics, Graduate School of Medicine, Hokkaido University, Sapporo, Japan.} % to be revised

\email{yokotai@pop.med.hokudai.ac.jp} % to be revised

\begin{abstract}

Randomized controlled trials are often infeasible in rare diseases because of limited sample size, high costs, or lengthy durations.
Hybrid control designs address this by combining a concurrent control arm with historical data.
However, existing methods such as propensity score matching (PSM) borrow historical data only at the analysis stage, which limits power gains when covariate distributions differ between historical and concurrent data.

We propose a stratified block randomization method for hybrid control trials.
Within each stratum, more concurrent subjects are allocated to the treatment arm than to the concurrent control arm, and the difference is filled with historical controls so that the treatment and hybrid control arms are balanced 1:1; once the available historical controls in the stratum are used or a pre-specified overall cap is reached, allocation returns to 1:1.
Historical controls are incorporated either by sampling or by weighting.
We compared both methods with PSM in terms of type I error rate, power, and sample size via Monte Carlo simulations.

The sampling method maintained the type I error rates when there was neither heterogeneity in the outcome model nor unmeasured covariates.
It also had power comparable to or greater than that of PSM, particularly when historical controls were unevenly distributed across strata.
The power of the weighting method increased with the pooled historical sample size, but its type I error rate was inflated when covariates beyond the stratification factors were present or when stratification factors were dichotomized from continuous covariates.

The proposed method incorporates historical data at the allocation stage rather than only at the analysis stage, enabling more efficient use of historical data.

\end{abstract}

\keywords{Hybrid control, Randomization, Historical control, Clinical trials}

\maketitle

\section{Introduction}

% Establish: hybrid control designs and the exchangeability challenge
Randomised controlled trials (RCTs) evaluate a new treatment against a control arm.
Recruiting subjects can be slow, costly, or infeasible when the eligible population is small, as in rare diseases or paediatric oncology.
Hybrid control designs augment the concurrent control arm with historical controls from previous trials or real-world data.\cite{pocock_combination_1976, damone_incorporating_2024, tan_augmenting_2022}
Such designs are supported by regulatory guidance and have been applied across therapeutic areas.\cite{jahanshahi_use_2021, li_hybrid-control_2022, sengupta_emulating_2023, van_eijk_hybrid_2023}
Since treatment assignment among concurrent subjects remains randomised, the comparison between concurrent subjects is unconfounded. The validity of the hybrid comparison, however, depends on the exchangeability of historical and concurrent controls. When baseline covariate distributions differ between the two cohorts, the estimated treatment effect can be biased.\cite{pocock_combination_1976}

% Limit: prior allocation-adaptive designs set an aggregate effective-sample-size ratio (Hobbs, Guo); the closest, covariate-local design (Jin) drives borrowing from outcomes
Existing designs typically borrow historical information at the analysis stage, for example through propensity score matching, which selects historical controls that resemble the concurrent cohort.\cite{austin_introduction_2011, lin_propensity_2018}
Matching yields approximately unbiased estimates when the propensity score model is correctly specified, but discards historical controls that lack a match, losing information the design could have used.
Allocation in these designs follows a fixed ratio that ignores the number of historical controls each stratum can supply.
When historical controls are unevenly distributed across strata, a fixed allocation ratio leaves the treatment and hybrid control arms imbalanced within strata, limiting the efficiency gained from borrowing.\cite{fu_covariate_2023, struebing_augmenting_2024}
Other designs adapt the allocation ratio using information from the historical data.
Some set a single trial-wide ratio from the effective sample size of the historical data, that is, the number of concurrent controls that would provide the same amount of information.\cite{hobbs_adaptive_2013, guo_adaptive_2024}
Like a fixed ratio, such a single ratio cannot balance the treatment and hybrid control arms within strata when historical controls are unevenly distributed, and reweighting the historical controls to correct the imbalance reduces their effective sample size.
Others adapt the ratio within each stratum according to the commensurability of historical and concurrent control outcomes, that is, how closely the two agree,\cite{jin_bayesian_2023} and thus depend on accumulated outcome data rather than baseline covariates alone.
Since that type of allocation depends on accumulated outcomes, it raises concerns about temporal trend bias and type I error rate associated with response-adaptive randomization.\cite{korn_outcome-adaptive_2011}

% Propose: stratum-aware allocation, design intent, and theoretical positioning
Therefore, we propose a stratified randomization procedure for hybrid control trials that incorporates the number of available historical controls in each stratum into the allocation.
When historical controls are available in a stratum, the procedure allocates more concurrent subjects to the treatment arm, so that the treatment arm and the hybrid control arm, which combines concurrent and borrowed historical controls, are balanced 1:1 within the stratum.
Allocation returns to a regular 1:1 ratio of the treatment and concurrent control arms once the number of borrowed historical control subjects in a stratum reaches the number available (the stratum quota), or the total number borrowed reaches a pre-specified maximum for the entire trial (the overall borrowing cap).
The quota in each stratum and the overall cap bound the number of historical controls that can be borrowed.
Since the allocation uses the per-stratum counts of historical controls, the design can attain a within-stratum balance that a fixed-ratio allocation with analysis-stage borrowing cannot provide.
The aim is to improve the precision of the treatment effect estimate for a given number of concurrent subjects; equivalently, since part of the control arm is supplied by historical controls, a target precision can be achieved with fewer concurrent controls.
Since allocation depends on strata and the availability of historical controls, but not on observed outcomes, the procedure is covariate-adaptive rather than response-adaptive.
The procedure, therefore, avoids the additional complications associated with outcome-dependent allocation, such as temporal-trend bias and inflation of the type I error rate.
The treatment effect can be estimated with a model-assisted estimator for which large-sample theory has been developed under covariate-adaptive randomization.\cite{Bugni2018_JASA, Ye2023_JASA, Bannick2025_Biometrika}

% ¶4 — Roadmap
The remainder of this paper describes the allocation algorithm, gives large-sample justification for the resulting estimator and test, and compares the design with standard stratified randomization and propensity-score-based borrowing in simulations.

\section{Proposed Method}

\subsection{Notation}

We consider a hybrid controlled clinical trial that combines a concurrent randomised cohort with a historical control dataset.
The trial randomises $n_T$ subjects to the treatment arm and $n_C$ to the concurrent control arm.
For each subject, $A_i \in \{0,1\}$ denotes the treatment indicator (with $A_i = 0$ for the control and $A_i = 1$ for the treatment) and $Y_i$ the observed binary or continuous outcome.
Every subject belongs, by observed baseline factors, to one of $K$ strata, and $k_i \in \{1, \ldots, K\}$ denotes the stratum of subject $i$.
The source indicator $S_i \in \{0,1\}$ distinguishes the two data sources, with $S_i = 1$ for a subject in the concurrent randomised cohort (treatment or control arm) and $S_i = 0$ for a subject in the historical control cohort.
The historical dataset provides $h_k$ candidate control subjects in stratum $k$, which we call the stratum quota. 
During the trial, $b_k$ of these ($b_k \le h_k$) are borrowed into the hybrid control arm, giving $b = \sum_{k=1}^{K} b_k$ borrowed controls in total, which is bounded by a pre-specified trial-wide cap $b^{\max}$.

\subsection{Design overview}

The proposed design determines the allocation within each stratum from the available size of historical controls.
In a stratum with available historical controls, each randomization block assigns more concurrent subjects to the treatment arm than to the concurrent control arm, and borrows as many historical controls into the hybrid control arm as the difference between the two, so that the treatment and hybrid control arms are balanced.
Once $b_k$ reaches the stratum quota $h_k$, or $b$ reaches the trial-wide cap $b^{\max}$, allocation in that stratum returns to balanced blocks, i.e. balanced between the concurrent treatment and control arms.
The borrowing is thus bounded separately within each stratum by the quota $h_k$ and jointly across strata by the cap $b^{\max}$.

\subsection{Allocation rule}

Within a stratum, borrowing is limited by the stratum quota $h_k$ and by the trial-wide cap $b^{\max}$.

We call a block with $r$ treatment and $s$ concurrent control slots, where $r > s \ge 1$, a full historically-linked block.
Initially, each such block borrows $r-s$ historical controls into the hybrid control arm.
Provided that both $b_k < h_k$ and $b < b^{\max}$, concurrent subjects in stratum $k$ are randomised in historically-linked blocks.
Before a block is generated, set
\begin{equation}\label{eq:excess}
  e = \min\{\, r - s,\ h_k - b_k,\ b^{\max} - b \,\}.
\end{equation}
Then, the block is constructed with $s + e$ treatment and $s$ concurrent control slots, and $e$ historical controls are borrowed for it.
A block with $e = r - s$ is full.
A block with $1 \le e < r - s$ arises only when the remaining stratum quota $h_k - b_k$ or the remaining trial-wide allowance $b^{\max} - b$ is smaller than $r - s$; in that case $e$ equals the remaining quantity, so borrowing it exhausts the quota or the cap.
Such a block is therefore shortened and is necessarily the final historically-linked block of the stratum.
A completed block contributes $s + e$ subjects to the treatment arm and $s + e$ to the hybrid control arm ($s$ concurrent and $e$ historical), so the two arms are balanced within the stratum.
We refer to the point at which $b_k = h_k$ or $b = b^{\max}$ as the switch point of stratum $k$.
From then on, no controls can be borrowed ($e = 0$) and the remaining subjects in the stratum are randomised in balanced blocks.
By construction, $b_k$ never exceeds $h_k$ and $b$ never exceeds $b^{\max}$.
The block parameters $(r,s)$ may also vary across strata, using $(r_k, s_k)$ for stratum $k$, without changing the allocation rule.

\begin{algorithm}[H]
  \caption{Allocation with historical control borrowing}
  \label{alg:hybrid}
  \DontPrintSemicolon  
  \KwIn{stratum $k_i \in \{1,2,\cdots,K\}$ of each arriving subject $i$; full-block parameters $(r,s)$, $r > s \ge 1$; stratum quotas $h_k$, the number of historical controls available for borrowing in stratum $k$; trial-wide cap $b^{\max}$}  
  \KwOut{treatment indicators $\{A_i\}$ and borrowed counts $\{b_k\}$}
  Initialise $b_k \leftarrow 0$ and an empty current block $B_k$ for every stratum $k$, and $b \leftarrow 0$\;
  \For{each arriving concurrent subject $i = 1, 2, \ldots$}{
    $k \leftarrow k_i$\;
    \If{$B_k$ is empty}{
      $e \leftarrow \min\{\, r - s,\ h_k - b_k,\ b^{\max} - b \,\}$\;
      \eIf{$e \ge 1$}{
        $b_k \leftarrow b_k + e$,\quad $b \leftarrow b + e$\;
        $B_k \leftarrow$ random permutation of $s + e$ treatment and $s$ concurrent control slots\;
      }{
        $B_k \leftarrow$ random permutation of a balanced block\;
      }
    }
    $A_i \leftarrow$ next position of $B_k$, then remove it from $B_k$\;
  }
  \Return $\{A_i\}$, $\{b_k\}$\;
\end{algorithm}

\subsection{Testing and estimation}

Once allocation is complete, the number of historical controls to combine with the concurrent control arm in stratum $k$ equals $b_k$.
At the analysis stage, the hybrid control arm can be formed in either of two ways:

\begin{enumerate}[label=\arabic*)]
\item \textbf{Sampling.} 
Exactly $b_k$ subjects are selected by random sampling from the stratum-$k$ pool of $h_k$ historical control subjects.
When baseline covariates beyond the stratifying factors are expected to differ between the historical and concurrent cohorts, they may instead be selected by propensity score matching.\cite{lin_propensity_2018}

\item \textbf{Weighting.}
All $h_k$ historical control subjects in the stratum-$k$ pool are retained, each carrying the weight
\begin{equation}\label{eq:weight}
  w_k = \frac{b_k}{h_k}, \qquad \forall k, \; 0 \le w_k \le 1,
\end{equation}
so that the total weight contributed by these $h_k$ subjects equals $b_k$.
If additional covariates are a concern, the propensity score can be used to restrict the historical pool before weighting.
\end{enumerate}

Both constructions can be written with a subject-level analysis weight $\omega_i$, with $\omega_i = 1$ for concurrent subjects and $\omega_i = w_k$ for historical control subjects in stratum $k$.
Let $\mathcal{T}_k$, $\mathcal{C}_k$, and $\mathcal{H}_k$ be the index sets of the treated, concurrent control, and historical control subjects in stratum $k$.
Under sampling, $\mathcal{H}_k$ consists of the $b_k$ selected historical controls and $w_k = 1$.
Under weighting, $\mathcal{H}_k$ consists of all $h_k$ historical controls and $w_k = b_k/h_k$.
In either case, $\sum_{i \in \mathcal{H}_k} \omega_i = b_k$.
Under sampling, the hybrid control arm consists of unweighted subjects, so the trial can be analysed in the usual way, for example by a linear regression of the outcome on the treatment indicator and the stratification factors.
Under weighting, the historical controls carry weights $w_k < 1$, and the analysis must account for them.
We therefore derive the estimator and its variance for the general weighted case below; setting $w_k = 1$ recovers the sampling construction.

\noindent\textit{Estimand and estimator.} 
Let $Y_i(a)$, $a \in \{0,1\}$, be the potential binary or continuous outcomes of subject $i$ under control ($a=0$) and treatment ($a=1$), with $Y_i = A_i Y_i(1) + (1 - A_i) Y_i(0)$ for concurrent subjects.
Let $\mu_a(k) = E[Y(a) \mid k_i = k, S_i = 1]$ and $\sigma_a^2(k) = \mathrm{Var}(Y(a) \mid k_i = k, S_i = 1)$ denote the outcome mean and variance in each concurrent arm of stratum $k$, and let $\sigma_{0H}^2(k)$ denote the outcome variance of the historical control arm in stratum $k$.
We allow the historical and concurrent control variances to differ, that is, we do not assume $\sigma_{0H}^2(k) = \sigma_0^2(k)$.
Let $p_k$ be the proportion of concurrent subjects belonging to stratum $k$ in the concurrent population.
The treatment effect in the concurrent population is
\begin{equation}\label{eq:estimand}
  \theta = E[Y(1) - Y(0) \mid S = 1] = \sum_{k=1}^{K} p_k\,\delta_k, \qquad \delta_k = \mu_1(k) - \mu_0(k).
\end{equation}
Let $n_{T,k} = |\mathcal{T}_k|$, $n_{C,k} = |\mathcal{C}_k|$, and $n = n_T + n_C$, and write $n_{1,k} = n_{T,k}$ and $n_{0,k} = n_{C,k} + b_k$ for the sizes of the treatment arm and the hybrid control arm in stratum $k$.
We estimate the arm means by
\begin{equation}\label{eq:armmeans}
  \hat\mu_1(k) = \frac{1}{n_{1,k}} \sum_{i \in \mathcal{T}_k} Y_i, \qquad
  \hat\mu_0(k) = \frac{1}{n_{0,k}} \Bigl( \sum_{i \in \mathcal{C}_k} Y_i + \sum_{i \in \mathcal{H}_k} \omega_i Y_i \Bigr),
\end{equation}
and the treatment effect by
\begin{equation}\label{eq:thetahat}
  \hat\theta = \sum_{k=1}^{K} \hat p_k \hat\delta_k, \qquad \hat\delta_k = \hat\mu_1(k) - \hat\mu_0(k), \qquad \hat p_k = \frac{n_{T,k} + n_{C,k}}{n}.
\end{equation}
The standardising proportions $\hat p_k$ are computed from concurrent subjects only, so the borrowed controls do not shift the target population.
Estimator \eqref{eq:thetahat} is the covariate-adjusted estimator with a working model saturated in stratum and arm, which for concurrent data coincides with the augmented inverse propensity weighted (AIPW) estimator.\cite{Robins1994_JASA, Ye2023_JASA, Bannick2025_Biometrika}
Its consistency rests on the randomization and on the exchangeability of historical and concurrent controls in their expected outcomes within strata (Assumption (A3) in Section~\ref{sec:largesample}), not on the correctness of a working model.

\noindent\textit{Estimating equations.} 
We derive the variance of $\hat\theta$ by M-estimation. Let $\vartheta = (\{\mu_1(k)\}, \{\mu_0(k)\}, \{p_k\}, \theta)^\top$ and let $O_i = (Y_i, S_i, A_i, k_i, \omega_i)$ be the record of subject $i$. The estimating function $\psi(O_i; \vartheta)$ stacks
\begin{align}
  \psi_{\mu_1,k}(O_i) &= S_i A_i \mathbbm{1}(k_i = k)\,\{Y_i - \mu_1(k)\}, \label{eq:psi1}\\
  \psi_{\mu_0,k}(O_i) &= \mathbbm{1}(k_i = k)\,\{(1 - A_i) S_i + \omega_i (1 - S_i)\}\,\{Y_i - \mu_0(k)\}, \label{eq:psi0}\\
  \psi_{p,k}(O_i) &= S_i \{\mathbbm{1}(k_i = k) - p_k\}, \label{eq:psip}\\
  \psi_{\theta}(O_i) &= \theta - \textstyle\sum_{k} p_k \{\mu_1(k) - \mu_0(k)\}. \label{eq:psith}
\end{align}
Solving $\sum_i \psi_{\mu_1,k}(O_i) = 0$, $\sum_i \psi_{\mu_0,k}(O_i) = 0$, and $\sum_i \psi_{p,k}(O_i) = 0$ yields the arm means in \eqref{eq:armmeans} and the proportions $\hat p_k$.
The last component $\psi_\theta$ involves no data; it defines $\theta$ as the $p_k$-weighted average of the stratum-specific effects, so solving it gives $\hat\theta$ in \eqref{eq:thetahat}.

\noindent\textit{Sandwich variance and influence function.} 
By the general theory of M-estimation, the asymptotic variance of $\hat\vartheta$ is $\mathbf{B}^{-1}\mathbf{M}(\mathbf{B}^{-1})^\top$ with bread $\mathbf{B} = E[-\partial\psi/\partial\vartheta^\top]$ and meat $\mathbf{M} = E[\psi\psi^\top]$, and the influence function is $\mathrm{IC}(O_i) = \mathbf{B}^{-1}\psi(O_i; \vartheta)$.
With $\pi_k = n_{1,k}/(n_{T,k} + n_{C,k})$, the non-zero elements of $\mathbf{B}$ are
\begin{gather}
  B_{\mu_1(k),\mu_1(k)} = p_k \pi_k, \qquad B_{\mu_0(k),\mu_0(k)} = \frac{n_{0,k}}{n}, \qquad B_{p_k,p_k} = 1, \label{eq:bread1}\\
  B_{\theta,\mu_1(k)} = p_k, \qquad B_{\theta,\mu_0(k)} = -p_k, \qquad B_{\theta,p_k} = \delta_k, \qquad B_{\theta,\theta} = -1, \label{eq:bread2}
\end{gather}
so that the $\theta$ row of $\mathbf{B}^{-1}$ has elements $1/\pi_k$, $-n p_k/n_{0,k}$, $\delta_k$, and $-1$.
Since $\psi_\theta$ contains no data, the influence function of $\hat\theta$ is
\begin{equation}\label{eq:ic}
  \mathrm{IC}_{\hat\theta}(O_i) = \sum_{k=1}^{K} \Bigl[ \frac{1}{\hat\pi_k} \psi_{\mu_1,k}(O_i) - \frac{n \hat p_k}{n_{0,k}} \psi_{\mu_0,k}(O_i) + \hat\delta_k \psi_{p,k}(O_i) \Bigr].
\end{equation}
With residuals $\hat e_i = Y_i - \hat\mu_{A_i}(k_i)$, where historical controls take $A_i = 0$, the influence function for a subject in stratum $k$ is
\begin{equation}\label{eq:if}
  \mathrm{IC}_{\hat\theta}(O_i) =
  \begin{cases}
    \dfrac{1}{\hat\pi_k}\,\hat e_i + (\hat\delta_k - \hat\theta), & i \in \mathcal{T}_k, \\[2ex]
    -\dfrac{n \hat p_k}{n_{0,k}}\,\hat e_i + (\hat\delta_k - \hat\theta), & i \in \mathcal{C}_k, \\[2ex]
    -\dfrac{n \hat p_k}{n_{0,k}}\,w_k\,\hat e_i, & i \in \mathcal{H}_k.
  \end{cases}
\end{equation}
The first terms come from $\psi_{\mu_1,k}$ and $\psi_{\mu_0,k}$ and describe within-stratum variability.
The term $\hat\delta_k - \hat\theta$ comes from $\psi_{p,k}$ and describes between-stratum variability; it enters only for concurrent subjects, because historical subjects do not contribute to $\hat p_k$.

\noindent\textit{Variance estimator.} 
Summing the squared influence functions within each group and stratum and using $\mathrm{Var}(\hat\theta) = n^{-2} \sum_{i} \mathrm{IC}_{\hat\theta}(O_i)^2$ gives
\begin{equation}\label{eq:varsand}
\begin{split}
  \widehat{\mathrm{Var}}(\hat\theta) = {} & \sum_{k=1}^{K} \hat p_k^2 \biggl[ \frac{1}{n_{1,k}^2} \sum_{i \in \mathcal{T}_k} \hat e_i^2 + \frac{1}{n_{0,k}^2} \Bigl\{ \sum_{i \in \mathcal{C}_k} \hat e_i^2 + w_k^2 \sum_{i \in \mathcal{H}_k} \hat e_i^2 \Bigr\} \biggr] \\
  & + \frac{1}{n} \sum_{k=1}^{K} \hat p_k (\hat\delta_k - \hat\theta)^2.
\end{split}
\end{equation}
The first term is the within-stratum variance of the arm means, and the second is the between-stratum variance of the stratum-specific effects.
Borrowing enters the first term through the weighted historical residuals, with total weight $b_k$ and sum of squared weights $h_k w_k^2$.
Since each squared residual estimates the variance of its own observation, Equation \eqref{eq:varsand} does not require $\sigma_{0H}^2(k) = \sigma_0^2(k)$.

To correct the downward bias of Equation \eqref{eq:varsand} in small samples, we apply the HC3 correction $\hat e_i^2/(1 - \ell_i)^2$. For the arm-by-stratum means, the leverages are
\begin{equation}\label{eq:leverage}
  \ell_{1,k} = \frac{1}{n_{1,k}}, \qquad \ell_{0,k} = \frac{1}{n_{0,k}}, \qquad \ell_{H,k} = \frac{w_k}{n_{0,k}}
\end{equation}
for treated, concurrent control, and historical control subjects, respectively. The resulting variance estimator is
\begin{equation}\label{eq:varhat}
\begin{split}
  \widehat{\mathrm{Var}}_{\mathrm{HC3}}(\hat\theta) = {} & \sum_{k=1}^{K} \hat p_k^2 \biggl[ \frac{1}{n_{1,k}^2} \sum_{i \in \mathcal{T}_k} \frac{\hat e_i^2}{(1 - \ell_{1,k})^2} \\
  & + \frac{1}{n_{0,k}^2} \Bigl\{ \sum_{i \in \mathcal{C}_k} \frac{\hat e_i^2}{(1 - \ell_{0,k})^2} + w_k^2 \sum_{i \in \mathcal{H}_k} \frac{\hat e_i^2}{(1 - \ell_{H,k})^2} \Bigr\} \biggr] \\
  & + \frac{1}{n} \sum_{k=1}^{K} \hat p_k (\hat\delta_k - \hat\theta)^2.
\end{split}
\end{equation}

The confirmatory analysis tests $H_0: \theta = 0$ with the Wald statistic $\hat\theta / \widehat{\mathrm{se}}(\hat\theta)$, where $\widehat{\mathrm{se}}(\hat\theta) = \widehat{\mathrm{Var}}_{\mathrm{HC3}}(\hat\theta)^{1/2}$. The corresponding two-sided $(1-\alpha)$ confidence interval is $\hat\theta \pm z_{1-\alpha/2}\,\widehat{\mathrm{se}}(\hat\theta)$. The test attains its nominal level asymptotically (cf. Lemma~\ref{lem:level} in a later section).

The sampling construction is simpler than the weighting construction, since the hybrid control arm then $w_i=1$ subjects, and matches current propensity-score practice.
The weighting construction retains all historical controls and has a smaller sum of squared weights, though its advantage, like that of any borrowing, depends on how closely the historical and concurrent controls agree within strata.

\subsection{Illustrative example}

We illustrate the design with a fictive trial in Table~\ref{tab:allocation_example}.
The trial enrols 120 concurrent subjects, and 46 historical control subjects are available, with a borrowing cap of $b^{\max} = 40$.
Each historically-linked block uses $(r,s) = (4,1)$, assigning four treatment and one concurrent control slot and borrowing $r - s = 3$ historical controls into the hybrid control arm in full historically-linked blocks.
For comparison, the table also shows the allocation under a conventional method that fixes a 2:1 allocation ratio across all strata and selects historical controls by propensity score matching.
The conventional method aims for a hybrid control arm of 80 by randomising 40 concurrent controls and adding 40 historical controls, but it selects the historical controls at the analysis stage, regardless of the strata in which they are needed.
Under the conventional method, no stratum has the treatment and hybrid control arms balanced 1:1.
In particular, the hybrid control arm is considerably larger than the treatment arm in Stratum 2 (43 vs 33), whereas in Stratum 4, which has no historical controls, the treatment arm remains twice as large as the hybrid control arm (20 vs 10) under the 2:1 allocation.
In contrast, under the proposed design, the allocation ratio in each stratum follows the number of historical controls available, so that the treatment and hybrid control arms are balanced 1:1 in every stratum.

\begin{table}[h]
\small\sf\centering
\setlength{\tabcolsep}{3pt}
\caption{Example of allocation results by conventional and proposed methods}
\label{tab:allocation_example}
\resizebox{\textwidth}{!}{
\begin{tabular}{lrr rrr rrrr}
\toprule
 & \multicolumn{2}{c}{Data} & \multicolumn{3}{c}{Conventional method} & \multicolumn{4}{c}{Proposed method} \\
\cmidrule(lr){2-3} \cmidrule(lr){4-6} \cmidrule(lr){7-10}
 & & & & \multicolumn{2}{c}{Control} & & \multicolumn{2}{c}{Control} \\
\cmidrule(lr){5-6} \cmidrule(lr){8-9}
Stratum & Concurrent & Historical & Treatment & Concurrent & Historical & Treatment & Concurrent & Historical & Weight \\
\midrule
Stratum 1 &  10 & 10 &  7 &  3 &  8 &  8 &  2 &  6 &  6/10 \\
Stratum 2 &  50 & 30 & 33 & 17 & 26 & 39 & 11 & 28 & 28/30 \\
Stratum 3 &  30 &  6 & 20 & 10 &  6 & 18 & 12 &  6 &   6/6 \\
Stratum 4 &  30 &  0 & 20 & 10 &  0 & 15 & 15 &  0 &     - \\
\midrule
Total     & 120 & 46 & 80 & 40 & 40 & 80 & 40 & 40 & \\
\bottomrule
\end{tabular}
}
\smallskip
\begin{minipage}{\textwidth}
\footnotesize\raggedright
Conventional method: concurrent subjects randomised 2:1 (treatment:control) by stratified block randomization; historical controls selected by propensity score matching after all subjects are allocated. Proposed method: allocation by Algorithm~\ref{alg:hybrid} with $(r,s) = (4,1)$ and $b^{\max} = 40$; Historical gives the borrowed count $b_k$, and Weight gives $w_k = b_k/h_k$ under the weighting construction.
\end{minipage}
\end{table}

Table~\ref{tab:allocation_blocks} shows the block sequence for each stratum.
Since blocks are generated sequentially as subjects arrive, the sequence depends on the arrival order, and the table shows one possible realisation.
Stratum 1 holds $h_1 = 10$ historical controls, but all ten concurrent subjects are allocated in the first two full historically-linked blocks, so six are borrowed and four remain unused.
Stratum 3 holds $h_3 = 6$ historical controls and reaches its quota after two full historically-linked blocks, then switches to balanced blocks.
Stratum 4, which has no historical controls, uses balanced blocks throughout.
Stratum 2 is the only stratum affected by the trial-wide cap.
It uses nine full historically-linked blocks, borrowing 27 controls, followed by a shortened block of size three that borrows one more.
At that point, the trial-wide count reaches the cap $b^{\max} = 40$, and Stratum 2 switches to balanced blocks, having borrowed 28 rather than all $h_2 = 30$ of its controls.
Since only two concurrent subjects remain in Stratum 2, only the first two slots of the following balanced block are used.

\begin{table}[h]
\small\sf\centering
\setlength{\tabcolsep}{3pt}
\caption{Example of block sequences by stratum under the proposed method}
\label{tab:allocation_blocks}
\resizebox{\textwidth}{!}{
\begin{tabular}{lll}
\hline
Stratum & Block type & Allocation blocks \\
\hline
{Stratum 1} & Historically-linked & \{T, C, T, T, T\}, \{T, T, T, T, C\} \\
            & Balanced & --- \\
\hline
{Stratum 2} & Historically-linked & \{T, T, T, C, T\}, \{T, T, C, T, T\}, \{T, T, T, T, C\}, \{T, T, T, T, C\}, \\
            &        & \{T, T, T, C, T\}, \{T, T, C, T, T\}, \{T, T, T, T, C\}, \{T, T, T, T, C\}, \\
            &        & \{T, C, T, T, T\}, \{T, C, T\} \\
            & Balanced & \{C, T, T, C\}$^{\dagger}$ \\
\hline
{Stratum 3} & Historically-linked & \{C, T, T, T, T\}, \{C, T, T, T, T\} \\
            & Balanced & \{C, T, C, T\}, \{T, C, T, C\}, $\cdots$ \\
\hline
{Stratum 4} & Historically-linked & --- \\
            & Balanced & \{T, C, T, C\}, \{C, T, T, C\}, $\cdots$ \\
\hline
\end{tabular}
}
\smallskip
\begin{minipage}{\textwidth}
\footnotesize\raggedright
T, treatment slot; C, concurrent control slot.
Each full historically-linked block with $(r,s) = (4,1)$ borrows $r - s = 3$ historical controls into the hybrid control arm; the shortened block \{T, C, T\} in Stratum 2 borrows one.
The order of slots within each block is randomly permuted, and the sequence shown is one realisation. $\cdots$ denotes further balanced blocks until all concurrent subjects in the stratum are allocated. \\
$^{\dagger}$Only the first two slots are used, as no further concurrent subjects arrive in Stratum 2.
\end{minipage}
\end{table}

\subsection{Large-sample properties}\label{sec:largesample}

We make the following assumptions. 
%Proofs of the results stated below are given in the Appendix.
\begin{itemize}
\item[(A1)] \textit{Independence.} Within each cohort, the subject records are independent and identically distributed.
\item[(A2)] \textit{Stratification consistency.} The strata are defined consistently in the concurrent and historical control cohorts.
\item[(A3)] \textit{Historical--concurrent exchangeability.} Within each stratum, control subjects in the concurrent and historical cohorts have the same expected outcome.
\end{itemize}

\noindent\textit{Design properties.} By construction, the treatment and hybrid control arms are balanced 1:1 within every stratum at each switch point, so the design achieves strong balance.\cite{Hu2012_AnnStat, Bugni2018_JASA} The borrowed count never exceeds the quota $h_k$ or the cap $b^{\max}$.

\noindent\textit{Inference conditions.} Allocation depends only on the stratum and not on the outcomes, so the assignment mechanism is outcome-blind. Under this condition the model-assisted estimator remains asymptotically valid even though covariate-adaptive randomization makes the assignments dependent.\cite{Bugni2018_JASA, Bannick2025_Biometrika}

Asymptotic results take the concurrent sample size $n_T + n_C$ to be large with the cap $b^{\max}$ fixed, so that borrowing is confined to a fixed, finite portion of an otherwise standard 1:1 stratified trial.\cite{Hu2012_AnnStat, Bugni2018_JASA, Bannick2025_Biometrika}

\begin{lemma}[Consistency]\label{lem:consistency}
Under (A1)--(A3) and outcome-blind allocation, the estimator \eqref{eq:thetahat} is consistent for $\theta$.
\end{lemma}

\begin{lemma}[Asymptotic level]\label{lem:level}
Under (A1)--(A3), the Wald test based on \eqref{eq:thetahat} and the robust variance estimator \eqref{eq:varhat} attains asymptotic level $\alpha$ under the proposed design. Strong balance renders the conventional variance estimator conservative; the robust estimator removes this conservativeness.
\end{lemma}

\begin{proposition}[Efficiency gain]\label{prop:efficiency}
Under (A1)--(A3) with $\sigma_{0H}^2(k) = \sigma_0^2(k)$, borrowing $b_k$ historical controls by sampling reduces the variance of the estimated control mean in stratum $k$ by the factor
\begin{equation}\label{eq:efficiency}
  \frac{n_{C,k}}{n_{C,k} + b_k},
\end{equation}
so that the effective control sample size increases from $n_{C,k}$ to $n_{C,k} + b_k$. Under weighting the variance is $\sigma_0^2(k)\{n_{C,k} + b_k^2/h_k\}/(n_{C,k} + b_k)^2$, which is no larger than under sampling.
\end{proposition}

\begin{remark}
In trials of the size considered here, the uncorrected sandwich estimator \eqref{eq:varsand} is biased downward and inflates the type I error rate. The HC3 correction in \eqref{eq:varhat} removes this bias but tends to be conservative, particularly under weighting, where the hybrid control arm has a small effective size and hence large leverage.
\end{remark}

\begin{remark}
Assumption (A3) is an identification condition, required only at the level at which the analysis adjusts for the historical--concurrent difference. Measured within-stratum differences can be addressed by restricting the historical pool through propensity score matching, or by including additional confounders in the working model. Such adjustment replaces (A3) with the analogous assumption conditional on those confounders, which may hold when the stratum-level version does not. Unmeasured differences, or modification of the treatment effect by cohort, are removed neither by the design nor by the estimator, and they bias both point and interval estimates. This residual risk motivates a conservative choice of the cap $b^{\max}$.
\end{remark}

\section{Simulation Study}

\subsection{Simulation settings}

% ##### paragraph 1 ##### %
% ##############################
% purpose of simulation
% data generation process
% ##############################

We conducted a simulation study to evaluate the performance of the proposed method and compare it with conventional approaches.
Unless otherwise specified, the following data generating process was used throughout the simulations.
We generated a concurrent cohort ($ S_i = 1 $) and a historical cohort ($ S_i = 0 $).
The pooled historical sample size, that is, the number of historical subjects available to the trial, was $ n_H = \sum_{k} h_k \in \{ 40, 50, 60, 70, 80 \} $, drawn at random from the historical cohort.
For each subject $ i $, three independent covariates were drawn as $ L_{1i}, L_{2i}, L_{3i} \sim \text{Bernoulli}(0.5) $.
Let $ \mathbf{L}_i = ( L_{1i}, L_{2i}, L_{3i} )^\top $ denote the covariate vector.
These three binary covariates define $ K = 2^3 = 8 $ allocation strata.
The continuous outcome was generated from
\begin{equation}\label{eq:dgp}
Y_i = \text{MD} \cdot A_i + \boldsymbol{\beta}^\top \mathbf{L}_i + \varepsilon_i, \qquad \varepsilon_i \sim N(0, 2^2),
\end{equation}
where MD is the true mean difference, set to 0.9.
With this effect size and error variance, a concurrent 1:1 randomised trial without borrowing requires $n = 160$ subjects (80 per arm) for a power of approximately 80\%; this sample size is used for the benchmark below.
We set the regression coefficients for the covariates to $\boldsymbol{\beta} = (\beta_1, \beta_2, \beta_3)^\top = (-\log(3), -\log(2), \log(3))^\top$.

% ##### paragraph 2 ##### %
% ##############################
% explain scenarios
% what paramater are changed
% ##############################

We considered seven scenarios to evaluate the performance of the proposed method under various conditions, as summarized in the following description and Table \ref{tab:scenarios}.
Scenario 1 serves as a reference, and Scenarios 2--7 were each designed to reflect either heterogeneity or differences in covariate structure between concurrent and historical data.

\begin{description}[style=nextline]

    \item[Scenario 1: Ideal alignment] 
    There was no heterogeneity between concurrent and historical data. This was the simplest scenario.
    The data generating process was identical for both cohorts.

    \item[Scenario 2: Lack of strata support (historical)]
    We assumed different distributions of $\mathbf{L}_i$ between concurrent and historical data.
    In the historical data, $L_{1i}$ was fixed at 0, so that the four strata with $L_1 = 1$ contained no historical subjects.
    
    \item[Scenario 3: Lack of strata support (concurrent)]
    This scenario mirrors Scenario 2 with the roles of the two cohorts reversed: in the concurrent data, $L_{1i}$ was fixed at 0, so that the four strata with $L_1 = 1$ contained no concurrent subjects, while the historical data covered all eight strata.

    \item[Scenario 4: Measured covariate shift]
    We added two covariates $ \mathbf{L}'_i = ( L_{4i}, L_{5i} )^\top $, where
    \begin{align}
    ( L_{4i} \mid S_i ) &\sim \begin{cases}
    \text{Bernoulli}(0.5) & S_i = 1 \\
    \text{Bernoulli}(0.2) & S_i = 0
    \end{cases}, \\
    ( L_{5i} \mid S_i ) &\sim \begin{cases}
    \text{Bernoulli}(0.5) & S_i = 1 \\
    \text{Bernoulli}(0.8) & S_i = 0
    \end{cases}.
    \end{align}
    Let $ \boldsymbol{\beta}' = ( \log(1.7), -\log(1.5) )^\top $ denote the corresponding regression coefficients vector.
    We generated the outcome by adding $ \boldsymbol{\beta}'^\top\mathbf{L}'_i $ to Equation \eqref{eq:dgp}.
    
    \item[Scenario 5: Unmeasured covariate shift]
    The data generating process was identical to Scenario 4, but we treated $ \mathbf{L}'_i $ as unmeasured covariates in this scenario.
    
    \item[Scenario 6: Effect modification heterogeneity]
    This scenario introduced heterogeneity in the regression coefficients between concurrent and historical data.
    For historical data, the regression coefficients were set to $ \boldsymbol{\beta} = 0.5 \cdot ( -\log(3), -\log(2), \log(3) )^\top $.
    
    \item[Scenario 7: Dichotomized stratification]
    In this scenario, $ L'_{2i} $ followed a categorical distribution and $ L'_{3i} $ followed a normal distribution:
    \begin{align}
    P(L'_{2i} = l \mid S_i) &= \begin{cases}
    0.25 & l \in \{0.2, 0.4, 0.6, 0.8\}, \quad S_i = 1, \\ 
    0.4 & l \in \{0.2, 0.4\}, \quad S_i = 0, \\ 
    0.1 & l \in \{0.6, 0.8\}, \quad S_i = 0,
    \end{cases}
    \end{align}
    \begin{align}
    (L'_{3i} \mid S_i) &\sim \begin{cases}
    N(0, 1) & S_i = 1, \\
    N(0.8, 1) & S_i = 0,
    \end{cases}
    \end{align}
    \begin{align}
    L_{2i} = \mathbbm{1}(L'_{2i} \geq 0.5), \quad L_{3i} = \mathbbm{1}(L'_{3i} \geq 0.5).
    \end{align}
    The outcome was generated from Equation~\eqref{eq:dgp} with $L'_{2i}$ and $L'_{3i}$ in place of $L_{2i}$ and $L_{3i}$, whereas the allocation strata and the stratified analysis used the dichotomised covariates $L_{2i}$ and $L_{3i}$.
    As a result, differences in $L'_{2i}$ and $L'_{3i}$ between the cohorts remained within strata.

\end{description}

\begin{table}[h]
\small\sf\centering
\caption{Simulation scenarios}
\label{tab:scenarios}
\resizebox{\textwidth}{!}{
\begin{tabular}{lcccc}
\hline
No & Scenario & Strata dist. differs & Covariate shift & Effect modification differs \\
\hline
1 & Ideal alignment & No & No & No \\
2 & Lack of strata support (historical) & Yes & No & No \\
3 & Lack of strata support (concurrent) & Yes & No & No \\
4 & Measured covariate shift & No & \makecell{Yes \\ (measured)} & No \\
5 & Unmeasured covariate shift & No & \makecell{Yes \\ (unmeasured)} & No \\
6 & Effect modification heterogeneity & No & No & Yes \\
7 & Dichotomised stratification & Yes & \makecell{Yes \\ (within strata)} & No \\
\hline
\end{tabular}
}
\smallskip
\begin{minipage}{\textwidth}
\footnotesize\raggedright
Strata dist. differs: the distribution of subjects across the allocation strata differs between the concurrent and historical cohorts.
Covariate shift: covariates other than the stratification factors differ in distribution between the cohorts.
Effect modification differs: the covariate effects on the outcome differ between the cohorts.
\end{minipage}
\end{table}

% ##### paragraph 3 ##### %
% ##############################
% compare methods
% what methods are changed
% why these methods are used
% ##############################

We evaluated and compared the performance of five methods.
The benchmark represents the ideal situation in which $n = 160$ concurrent subjects can be enrolled.
The other four methods assume that only $n = 120$ concurrent subjects are available and differ in whether and how historical controls are used.

\begin{description}[style=nextline]

    \item[1. Full concurrent RCT (Benchmark)]
    A fully powered concurrent cohort of $ n = 160 $ subjects was allocated in a 1:1 ratio ($ n_T \approx 80, n_C \approx 80 $) by standard stratified randomization with blocks of size four consisting of two treatment and two control slots.
    No historical data were included in the analysis stage.
    This method serves as a benchmark representing the ideal situation in which sufficient concurrent data are available.

    \item[2. Concurrent-only RCT (RCT only)]
    The concurrent cohort was allocated in a 2:1 ratio ($ n_T \approx 80, n_C \approx 40 $) by standard stratified randomization with blocks of size six consisting of four treatment and two control slots.
    No historical data were included in the analysis.
    This method allowed us to identify the contribution of historical data to the performance gains of PSM and the proposed method described below.

    \item[3. Propensity score matching (PSM)]
    We selected PSM as a comparison method because the proposed method operates within a frequentist framework, in which PSM is widely used for historical data.
    This method used the same allocation as RCT only but with historical data in the analysis stage.
    The propensity score for treatment exposure was estimated via logistic regression including all measured covariates ($ L_1 $--$ L_3 $, and also $ L_4 $ and $ L_5 $ in Scenario 4), and the logit of the propensity score was used for matching.
    We employed nearest-neighbour matching using the \texttt{MatchIt} package.
    The caliper width was set to 0.2 times the standard deviation of the logit of the propensity score, and was expanded iteratively until 40 matched historical subjects were selected.\cite{lin_propensity_2018}

    \item[4. Proposed method (Sampling)]
    The concurrent cohort was allocated by Algorithm~\ref{alg:hybrid} with full historically-linked blocks $ (r, s) = (4, 1) $, that is, four treatment and one concurrent control slot, borrowing $ r - s = 3 $ historical controls per block.
    The quota $ h_k $ was the pooled historical sample size in stratum $ k $, and the trial-wide cap was set to $ b^{\max} = 40 $.
    Once $ b_k = h_k $ or $ b = b^{\max} $, the stratum was randomised in balanced blocks of size four with two treatment and two concurrent control slots.
    At the analysis stage, $ b_k $ historical controls were sampled at random from the stratum-$ k $ pool.

    \item[5. Proposed method (Weighting)]
    The allocation was identical to that of the proposed method (sampling).
    At the analysis stage, all $ h_k $ historical controls in stratum $ k $ were retained with weight $ w_k = b_k / h_k $ from Equation \eqref{eq:weight}.
    
\end{description}

% ##### paragraph 4 ##### %
% ##############################
% performance metrics and details
% ##############################

The sampling, PSM, RCT only, and Benchmark analyses used a linear regression model including the treatment indicator and all measured covariates ($ L_1 $--$ L_3 $, and also $ L_4 $ and $ L_5 $ in Scenario 4), and tested the treatment coefficient with the Wald test using the model-based standard error.
The weighting analysis used the stratified estimator from Equation \eqref{eq:thetahat} with the weights as given by Equation \eqref{eq:weight} and the robust variance estimator \eqref{eq:varhat}; the derived variance was used only for the weighting analysis.
Since this analysis is stratified, it used only the three stratification factors $ L_1 $--$ L_3 $.
All tests were two-sided at a significance level of $ \alpha = 0.05 $.
We compared the methods by the type I error rate, the power, and the average sample size.
The type I error rate was evaluated using data in which only the treatment effect was set to the null ($ \text{MD} = 0 $), and the power using data with $ \text{MD} = 0.9 $.
Both were calculated as the proportion of simulations in which the null hypothesis was rejected.
The average sample size was summarised as the mean numbers of treated subjects $ n_T $, concurrent controls $ n_C $, and borrowed historical controls $ b $ over the simulations.
The type I error rate and power are shown for $ n_H = 40, 50, 60, 70, 80 $, and the average sample size for all values of $ n_H $.
We conducted 100,000 simulation iterations using R version 4.5.3.
With 100,000 iterations, the Monte Carlo standard error was approximately 0.07 percentage points for a type I error rate of 5\% and 0.13 percentage points for a power of 80\%.

\subsection{Results}

\begin{figure}[!htbp]
\centering
\includegraphics[width=\textwidth]{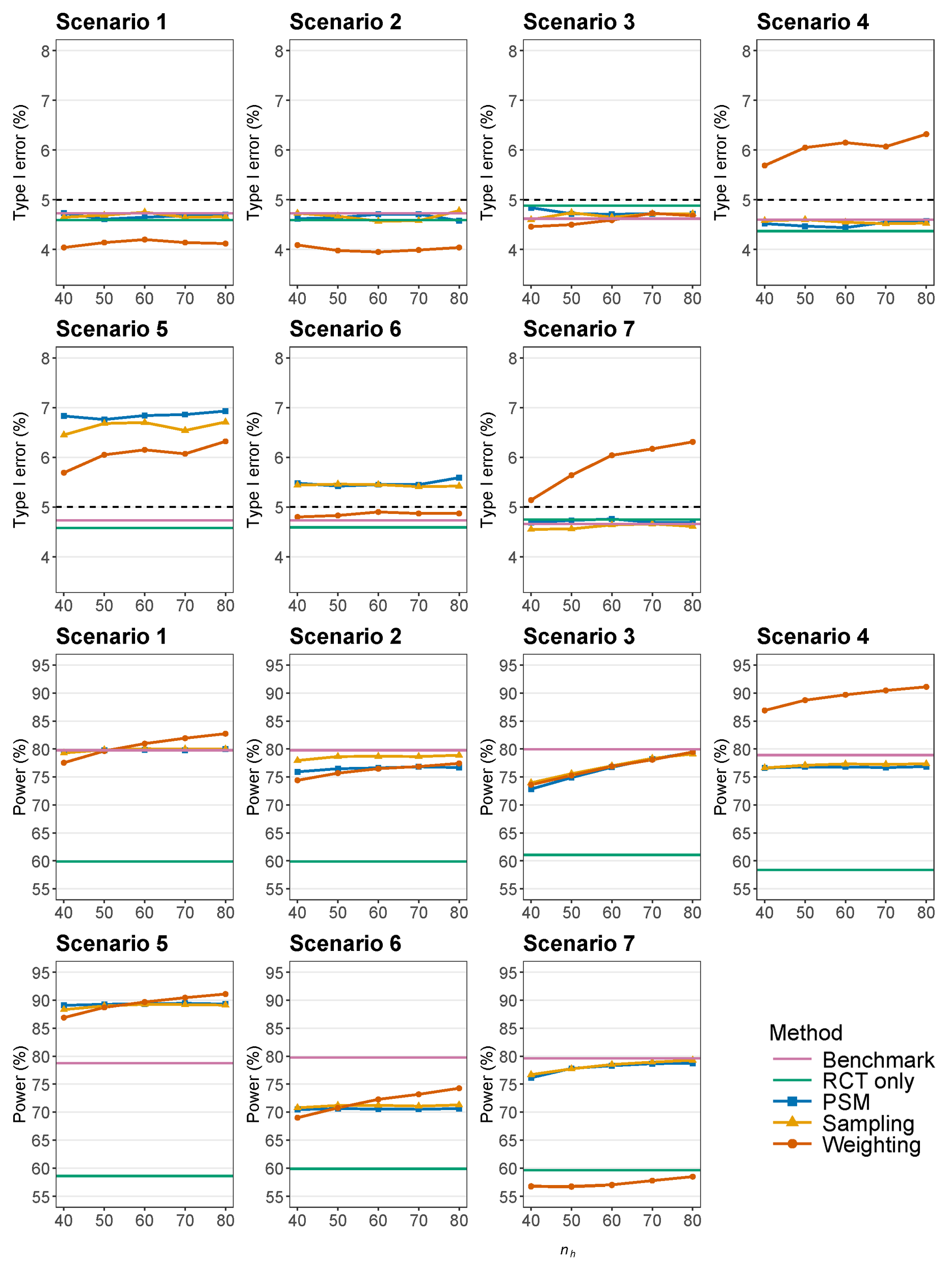}
\caption{Type I error rate and power by scenario and pooled historical sample size $ n_H $.}
\label{fig:alphapower}

\smallskip
\begin{minipage}{\textwidth}
\footnotesize\raggedright
$n_H = \sum_k h_k$, the pooled historical cohort size.
Benchmark, 1:1 stratified randomization with $n = 160$ concurrent subjects.
RCT only, 2:1 stratified randomization without historical controls ($n = 120$).
RCT only and Benchmark do not depend on $n_H$.
PSM, 2:1 stratified randomization with 40 historical controls selected by propensity score matching.
Sampling, proposed method with $b_k$ historical controls randomly sampled in each stratum.
Weighting, proposed method retaining all historical controls with weight $w_k = b_k/h_k$.
The dotted line indicates the nominal significance level of 5\%.
\end{minipage}
\end{figure}

\begin{table}[h]
\small\sf\centering
\setlength{\tabcolsep}{3pt}
\caption{Average sample size by scenario and pooled historical sample size $n_H$}
\label{tab:samplesize}
\resizebox{\textwidth}{!}{
\begin{tabular}{cccccc}
\toprule
Scenario & $n_H$ & Benchmark & RCT only & PSM & Sampling \& Weighting \\
\midrule
1 & 40 & \multirow{5}{*}{80.0 : 80.0 + 0.0} & \multirow{5}{*}{80.0 : 40.0 + 0.0} & 80.0 : 40.0 + 39.6 & 79.4 : 40.6 + 37.0 \\
  & 50 & & & 80.0 : 40.0 + 40.0 & 80.0 : 40.0 + 38.9 \\
  & 60 & & & 80.0 : 40.0 + 40.0 & 80.0 : 40.0 + 39.4 \\
  & 70 & & & 80.0 : 40.0 + 40.0 & 80.0 : 40.0 + 39.6 \\
  & 80 & & & 80.0 : 40.0 + 40.0 & 80.0 : 40.0 + 39.8 \\
\addlinespace
2 & 40 & \multirow{5}{*}{80.0 : 80.0 + 0.0} & \multirow{5}{*}{80.0 : 40.0 + 0.0} & 80.0 : 40.0 + 36.0 & 76.0 : 44.0 + 31.7 \\
  & 50 & & & 80.0 : 40.0 + 38.3 & 77.2 : 42.8 + 34.1 \\
  & 60 & & & 80.0 : 40.0 + 38.9 & 77.5 : 42.5 + 35.0 \\
  & 70 & & & 80.0 : 40.0 + 39.1 & 77.6 : 42.4 + 35.3 \\
  & 80 & & & 80.0 : 40.0 + 39.1 & 77.7 : 42.3 + 35.3 \\
\addlinespace
3 & 40 & \multirow{5}{*}{80.0 : 80.0 + 0.0} & \multirow{5}{*}{80.0 : 40.0 + 0.0} & 80.0 : 40.0 + 20.0 & 70.0 : 50.0 + 19.0 \\
  & 50 & & & 80.0 : 40.0 + 25.0 & 72.5 : 47.5 + 24.0 \\
  & 60 & & & 80.0 : 40.0 + 30.0 & 75.0 : 45.0 + 29.0 \\
  & 70 & & & 80.0 : 40.0 + 34.8 & 77.4 : 42.6 + 33.7 \\
  & 80 & & & 80.0 : 40.0 + 38.2 & 79.1 : 40.9 + 37.4 \\
\addlinespace
4 & 40 & \multirow{5}{*}{80.0 : 80.0 + 0.0} & \multirow{5}{*}{80.0 : 40.0 + 0.0} & 80.0 : 40.0 + 39.6 & 79.4 : 40.6 + 37.0 \\
  & 50 & & & 80.0 : 40.0 + 40.0 & 80.0 : 40.0 + 38.9 \\
  & 60 & & & 80.0 : 40.0 + 40.0 & 80.0 : 40.0 + 39.4 \\
  & 70 & & & 80.0 : 40.0 + 40.0 & 80.0 : 40.0 + 39.6 \\
  & 80 & & & 80.0 : 40.0 + 40.0 & 80.0 : 40.0 + 39.8 \\
\addlinespace
5 & 40 & \multirow{5}{*}{80.0 : 80.0 + 0.0} & \multirow{5}{*}{80.0 : 40.0 + 0.0} & 80.0 : 40.0 + 39.6 & 79.4 : 40.6 + 37.0 \\
  & 50 & & & 80.0 : 40.0 + 40.0 & 80.0 : 40.0 + 38.9 \\
  & 60 & & & 80.0 : 40.0 + 40.0 & 80.0 : 40.0 + 39.4 \\
  & 70 & & & 80.0 : 40.0 + 40.0 & 80.0 : 40.0 + 39.6 \\
  & 80 & & & 80.0 : 40.0 + 40.0 & 80.0 : 40.0 + 39.8 \\
\addlinespace
6 & 40 & \multirow{5}{*}{80.0 : 80.0 + 0.0} & \multirow{5}{*}{80.0 : 40.0 + 0.0} & 80.0 : 40.0 + 39.6 & 79.4 : 40.6 + 37.0 \\
  & 50 & & & 80.0 : 40.0 + 40.0 & 80.0 : 40.0 + 38.9 \\
  & 60 & & & 80.0 : 40.0 + 40.0 & 80.0 : 40.0 + 39.4 \\
  & 70 & & & 80.0 : 40.0 + 40.0 & 80.0 : 40.0 + 39.6 \\
  & 80 & & & 80.0 : 40.0 + 40.0 & 80.0 : 40.0 + 39.8 \\
\addlinespace
7 & 40 & \multirow{5}{*}{80.0 : 80.0 + 0.0} & \multirow{5}{*}{80.0 : 40.0 + 0.0} & 80.0 : 40.0 + 31.7 & 75.4 : 44.6 + 29.4 \\
  & 50 & & & 80.0 : 40.0 + 36.7 & 77.6 : 42.4 + 33.8 \\
  & 60 & & & 80.0 : 40.0 + 39.1 & 78.9 : 41.1 + 36.6 \\
  & 70 & & & 80.0 : 40.0 + 39.8 & 79.4 : 40.6 + 37.9 \\
  & 80 & & & 80.0 : 40.0 + 40.0 & 79.6 : 40.4 + 38.5 \\
\bottomrule
\end{tabular}
}

\smallskip
\begin{minipage}{\textwidth}
\footnotesize\raggedright
Entries are $n_T : n_C + b$, averaged over simulation iterations, where $n_T$, $n_C$, and $b$ are the numbers of treated subjects, concurrent controls, and borrowed historical controls, respectively.
Benchmark and RCT only do not depend on $n_H$. Sampling and Weighting share the same allocation and are therefore shown in a single column.
\end{minipage}
\end{table}

Figure~\ref{fig:alphapower} shows the type I error rate and power, and Table~\ref{tab:samplesize} shows the average sample size.

The sampling method had power comparable to or greater than that of PSM in all scenarios.
The gain was clearest in Scenarios 2 and 3, where the historical controls were unevenly distributed across strata.
In Scenario 2, the power of the sampling method was 77.9\% to 78.9\%, compared with 75.9\% to 76.8\% for PSM, a difference well beyond the Monte Carlo error.
In these scenarios, the proposed allocation kept the treatment and hybrid control arms balanced at 1:1 within each stratum, which led to higher efficiency.
The sampling method maintained the type I error rate near the nominal level (4.5\% to 4.8\%) in all scenarios except Scenario 5, with unmeasured confounding, and Scenario 6, with heterogeneity between the concurrent and historical data.
In these two scenarios, the type I error rate was inflated for both the sampling method and PSM (Scenario 5, 6.5\% to 6.9\%; Scenario 6, 5.4\% to 5.6\%).

The power of the weighting method increased with the pooled historical sample size, unlike the other methods.
In Scenario 1, it rose from 77.5\% at $ n_H = 40 $ to 82.7\% at $ n_H = 80 $, whereas the power of the sampling method and PSM stayed near 80\%.
However, since the weighting analysis is stratified, the type I error rate was inflated when covariates beyond the stratification factors were present or when a stratification factor was not derived from binary covariates.
In Scenario 4, the type I error rate of the weighting method was 5.7\% to 6.3\%, and in Scenario 7 it increased with $n_H$ from 5.1\%, within the Monte Carlo error of the nominal level, to 6.3\%.
In Scenario 7, the power of the weighting method was also lower than that of RCT only (56.8\% to 58.5\%).

\section{Discussion}

The power of the weighting method increased with the pooled historical sample size because its effective sample size increases.
Recall that $\omega_i = 1$ for concurrent controls and $\omega_i = w_k$ for historical controls in stratum $k$ under weighting.
Since $0 \le w_k \le 1$ implies $w_k^2 \le w_k$, the effective sample size of the hybrid control arm satisfies
\begin{equation}\label{eq:ess}
\begin{split}
  \mathrm{ESS}_{\mathrm{weighting}}
  &= \frac{\bigl(\sum_i \omega_i\bigr)^2}{\sum_i \omega_i^2}
  = \frac{\bigl(n_C + \sum_k h_k w_k\bigr)^2}{n_C + \sum_k h_k w_k^2} \\
  &\ge n_C + \sum_k h_k w_k = \mathrm{ESS}_{\mathrm{sampling}},
\end{split}
\end{equation}
where $\sum_k h_k w_k = b$ is the number of borrowed controls.
Under sampling, the effective sample size equals $n_C + b$ regardless of the pooled historical sample size.
Under weighting, $b$ is fixed by the allocation, but $\sum_k h_k w_k^2 = \sum_k b_k^2 / h_k$ decreases as $h_k$ grows, so the effective sample size increases.
This explains why only the weighting method gained power as $n_H$ grew.

Scenarios 2 and 3 suggest that balancing the treatment and hybrid control arms within each stratum improves efficiency.
In these scenarios, the historical controls were unevenly distributed across strata.
With a fixed 2:1 allocation, PSM could not match the borrowed controls to the strata in which they were needed.
The proposed allocation adapted the within-stratum ratio to the available historical controls and kept both arms at 1:1 within each stratum.

The weighting method benefits from its larger effective sample size, but it also loses information through dichotomization.
Since the analysis is stratified, covariates that are not stratification factors cannot be adjusted for, and a continuous covariate enters only through its dichotomised version.
In Scenarios 4 and 7, this residual imbalance within strata inflated the type I error rate, and in Scenario 7 it also reduced power.
The larger $n_H$ is, the more the analysis relies on the historical controls, and the more such residual differences matter.

In Scenarios 5 and 6, the type I error rate was inflated for all methods that borrowed historical controls, including PSM.
Unmeasured confounding and differences in covariate effects between the concurrent and historical data violate (A3), and neither the design nor the analysis can remove the resulting bias.
The cap $b^{\max}$ limits the amount of borrowing and therefore the size of this bias.

This study has several limitations.
First, the simulation considered a limited set of settings: a single concurrent sample size of 120, a borrowing cap of 40, block parameters $(r,s) = (4,1)$, one treatment effect, and three stratification factors with relatively large effects.
The relative performance of the methods may differ under other settings, for example with more strata, sparser strata, or smaller covariate effects.
Second, only a continuous outcome was evaluated, although the estimator and variance estimator apply equally to binary outcomes.
Third, the validity of borrowing relies on (A3), which cannot be verified from the data.
Finally, the choice of $b^{\max}$ was treated as given, and guidance on how to choose it in practice remains to be developed.

\section{Conclusion}

Balancing the treatment and hybrid control arms at 1:1 within each stratum, as in the proposed method, can improves efficiency.
Since the proposed method is a stratified design, a continuous allocation factor must be dichotomised, which loses information, and only a limited number of factors can be considered.
The development of designs that can use continuous covariates directly and account for many factors is a topic for future research.

\bibliographystyle{SageV}
\bibliography{references}

\end{document}